\documentclass[twocolumn,showpacs,preprintnumbers,amsmath,amssymb,a4paper,10pt]{revtex4}
\usepackage{amsthm}
\usepackage{mathrsfs}
\usepackage{graphicx}% Include figure files
\usepackage{dcolumn}% Align table columns on decimal point
\usepackage{bm}% bold math
\usepackage{amsmath}
\usepackage{subfigure}

\begin{document}

\title{Single-atom polarizer}

\author{Wen-An Li$^{1}$ and Lian-Fu Wei$^{1,2}$\footnote{E-mail: weilianfu@gmail.com}}

\address{1. State Key Laboratory of Optoelectronic Materials
and Technologies, School of Physics and Engineering, Sun Yat-Sen
University, Guangzhou 510275, China \\ 2. Quantum Optoelectronics Laboratory, School of Physics
and Technology, Southwest Jiaotong University, Chengdu 610031, China}

\begin{abstract}
Traditional polarizer provides a way to convert an unknown polarization into a specified polarization. According to Malus, the intensity of the transmission through such a polarizer is directly proportional to the square of the cosine of angle
between the transmission axes of the polarizer and the incident polarization. There, the intensity refers to the
collective behavior of {\it many} photons.
Here we propose a novel approach to realizing polarization-filtering at single-photon level.
We discuss how {\it a single} plane-polarized photon transports through a polarized
analyzer generated by a single atom (the so-called ``single-atom polarizer''), and provide a single quantum version of Malus's law.
We investigate the quantum scattering of a single photon by a controllable four-level atom inside a one-dimensional waveguide.
By using real-space theoretical approach, we obtain analytic expressions of the transmission spectrum of the photon.
Then, our numerical experiments show that the transmitted probability of the incident photon can be controlled by selecting the polarized-dependent transition configuration of the driven atom.
The application of such a single-atom polarizer to linear optical quantum information processing is also discussed.
\end{abstract}


\maketitle

%\emph{Introduction.--}
Polarization control of light is one of the main topics in optics~\cite{law}. According to the well-known Malus's law~\cite{book}, one can obtain certain polarized photons by placing a polarizer in front of an incident light beam. If $\alpha$ is the relative angle between the incoming wave and the orientation of the polarizer, Malus's law then says the intensity of the transmitted light is attenuated to be $I=I_0\cos^2 \alpha$ (with $I_0$ being the intensity of the incident beam). Classically, the intensity of a light beam is proportional to the square of electric amplitude and is really a collective behavior of many photons in the beam. The attenuation of the polarizer is due to the reflection and unavoidable absorption. For example,
around $38\%$ energy of the incident light is absorbed for the so-called Polaroid-type polarizers~\cite{p}. Therefore, photon loss should be a serious defect for the weak light through a polarizer based on the usual Malus's law. For example, when the detector shows no counts, it is not sure that the photon is reflected or absorbed by the polarizer. In fact, polarization-filtering at single-photon level is particularly important for the optical quantum communication and linear optical quantum computing~\cite{oqc1,oqc2}.

Recently, many efforts have been exerted to implement all-optical signal processing and all-optical quantum information processing~\cite{1,2}, in which one light signal controls another. Achieving this goal is a fundamental challenge because it requires an unique combination of large nonlinearities, low losses, as well as the light manipulation at single quantum level. The weak nonlinearities found in conventional media mean that large powers are required.
\begin{figure}[b]
\begin{center}
\includegraphics[width=0.38\textwidth]{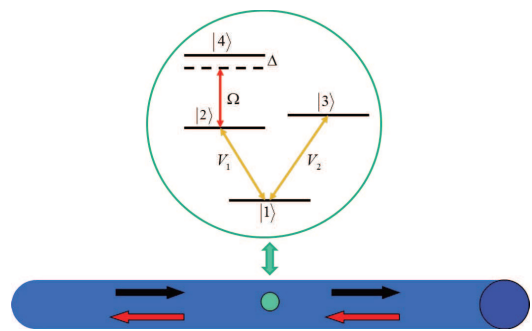}
\end{center}
\caption{Schematics of an ideal quantum waveguide
system: A single atom embedded in a one-dimensional quantum
waveguide. Here the four-level atom is indicated by a green
ball. The arrows denote the transporting directions of the photons, and $V_1$ ($V_2$)
represents the coupling strength between the atom and the photons. The transition $|2\rangle\leftrightarrow|4\rangle$ is driven by a classical laser with the coupling coefficient $\Omega$ and detuning $\Delta$.}
\end{figure}
Also, the single quantum operations imply that the single-photon transport along the optical waveguide and interaction with the other object should be well controlled. Indeed, recent studies of single photon scattered by a two or three-level atom inside a one-dimensional waveguide have revealed a series of interesting features~\cite{3,4,5,6,yan}. For example, Shen et al~\cite{3} have shown that an incident photon traveling along a one-dimensional waveguide would
be completely reflected, if the transition frequency of the two-level atom equals that of the transported
photon. This implies that a single two-level atom can be served as a mirror which completely reflects the resonant photon. Such model then was further investigated with three-level emitter, e.g., Zhang et al~\cite{6} showed that the photon can still be completely transmitted by the resonant-transition atom if another transition path of the atom is excited classically. Physically, in order to reach the one-dimensional regime the coupling between the photon and the transport mode in the waveguide has to be significantly strong (compared to transversal losses). Interestingly, great advances to reach such a regime have been reported recently in certain typical system; a line defect in photonic crystals~\cite{7,8} and superconducting transmission line~\cite{9,10,11}. With these experimental advances, various interesting transport properties of single photons, such as single-photon transistor~\cite{2}, frequency conversion~\cite{12}, and correlated photon transports~\cite{13}, etc. could be experimentally accessible.

In this paper we propose a so-called ``single-atom polarize'', generated by a driven four-level atom embedded in one dimensional waveguide. Similar to the usual polarizer, it allows the specific polarized photons pass (the photons with other polarizations are reflected without loss) one by one (i.e., at single-photon level) and thus is regarded as the quantum version of Malus' law. 

%\emph{Model.--}
The system considered here is sketched in Fig.~1, i.e., a four-level atom in an one-dimensional waveguide couples to the left- and right traveling-propagated modes of the waveguide. The transitions $|1\rangle\leftrightarrow|2\rangle$ and $|1\rangle\leftrightarrow|3\rangle$ are coupled to the photon mode with the strengths $V_1$ and $V_2$, respectively. Without loss of the generality, the transition $|2\rangle\leftrightarrow|4\rangle$ is driven by a classical laser with the coupling coefficient $\Omega$ and detuning $\Delta$.
The Hamiltonian of the system is given by
\begin{equation}
H=H_p+H_a+H_i
\end{equation}
where $H_p$ is the free photon Hamiltonian, $H_a$ is the free atomic Hamiltonian and $H_i$ describes the interaction between the photon and atom. In the real space coordinates $H_p$ reads
\begin{equation}
H_p=\int dx \left[-iv_gC_{R}^\dag(x)\partial_xC_{R}(x)+iv_gC_{L}^\dag(x)\partial_xC_{L}(x)\right],
\end{equation}
where $v_g$ denotes the group velocity of the photons. $C_{R}^\dag(x)$ $[C_{L}^\dag(x)]$ is the Fourier transform of the bosonic creation operator, describing the right(left)-traveling photon at position $x$ in the waveguide, respectively. The Hamiltonian of the driven atom $H_a$ is
\begin{eqnarray}
\nonumber  H_a &=& \left(\omega_2-i\frac{\gamma_2}{2}\right)a^\dag_2 a_2+\left(\omega_3-i\frac{\gamma_3}{2}\right)a^\dag_3 a_3 \\
   &+& \left(\omega_2+\Delta-i\frac{\gamma_4}{2}\right)a^\dag_4 a_4+\Omega(a^\dag_2 a_4+a_2 a^\dag_4)
\end{eqnarray}
with $a^\dag_i$ and $a_i$ being creation and annihilation operators, respectively. $a^\dag_ia_i=\hat{n}$ is the population operator for the atom at the state $|i\rangle$, and $\gamma_{i}(i=2,3,4)$ are the decay rates of corresponding excited states. Finally, the interaction term $H_i$
\begin{eqnarray}
\nonumber H_i&=&\int dx \delta(x)\left\{V_1\left[C_{R}^\dag(x)+C_{L}^\dag(x)\right]\sigma_{12}\right.\\
&+&\left.V_2\left[C_{R}^\dag(x)+C_{L}^\dag(x)\right]\sigma_{13}+\mathrm{h.c.}\right\}
\end{eqnarray}
describes the scattering between photons and the atom. Here, $\sigma_{ij}=a^\dag_ia_j$ is the transition operator from the state $|j\rangle$ to the state $|i\rangle$; $V_1$ and $V_2$ describe the interaction between each transition dipole moment and the photon field. $\delta(x)$ is introduced to describe the point-like interaction between the atom and photon occurring at $x=0$.
\begin{figure}[t]
\begin{center}
\subfigure[]{\includegraphics[width=0.35\textwidth]{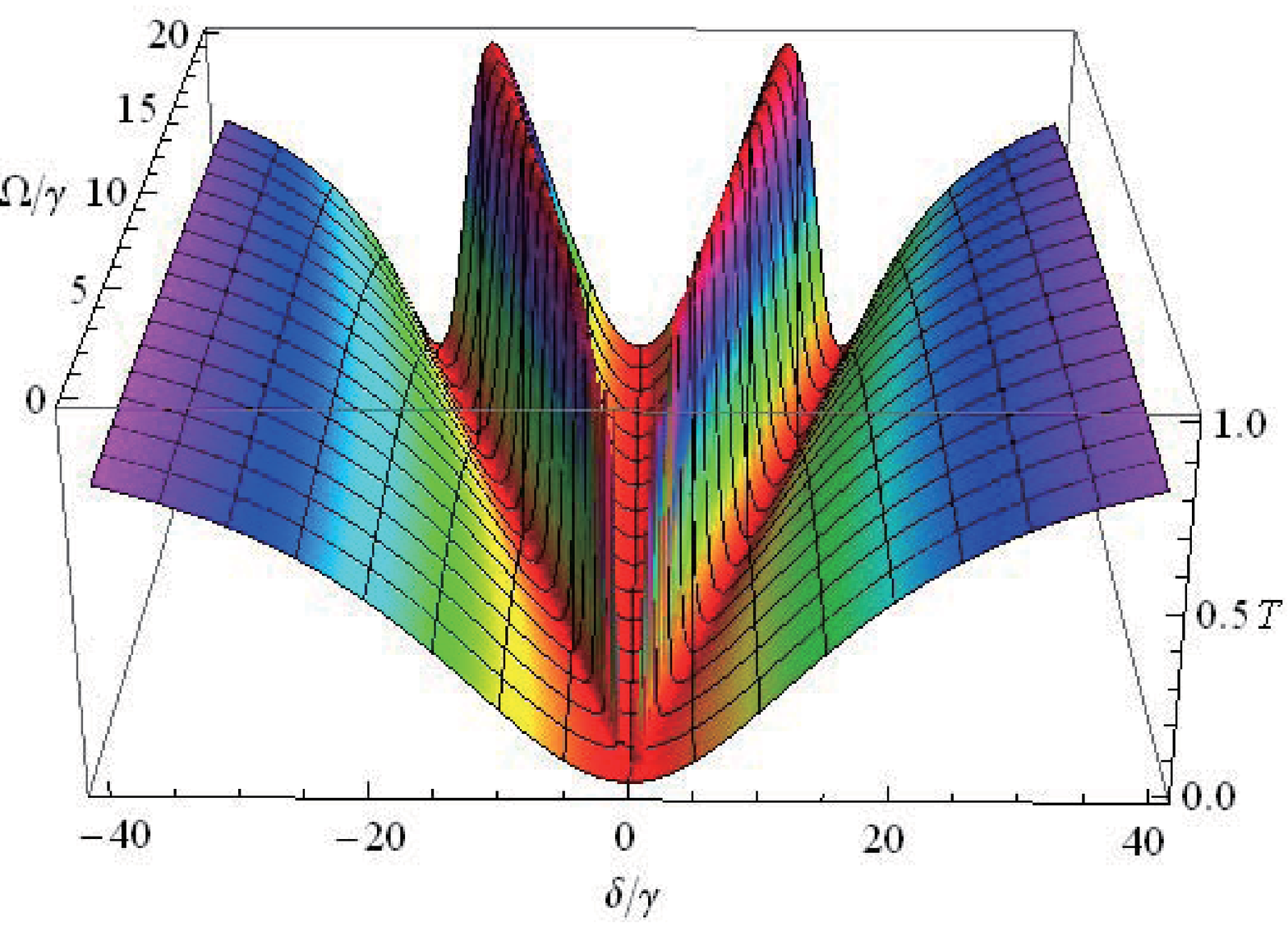}}
\subfigure[]{\includegraphics[width=0.35\textwidth]{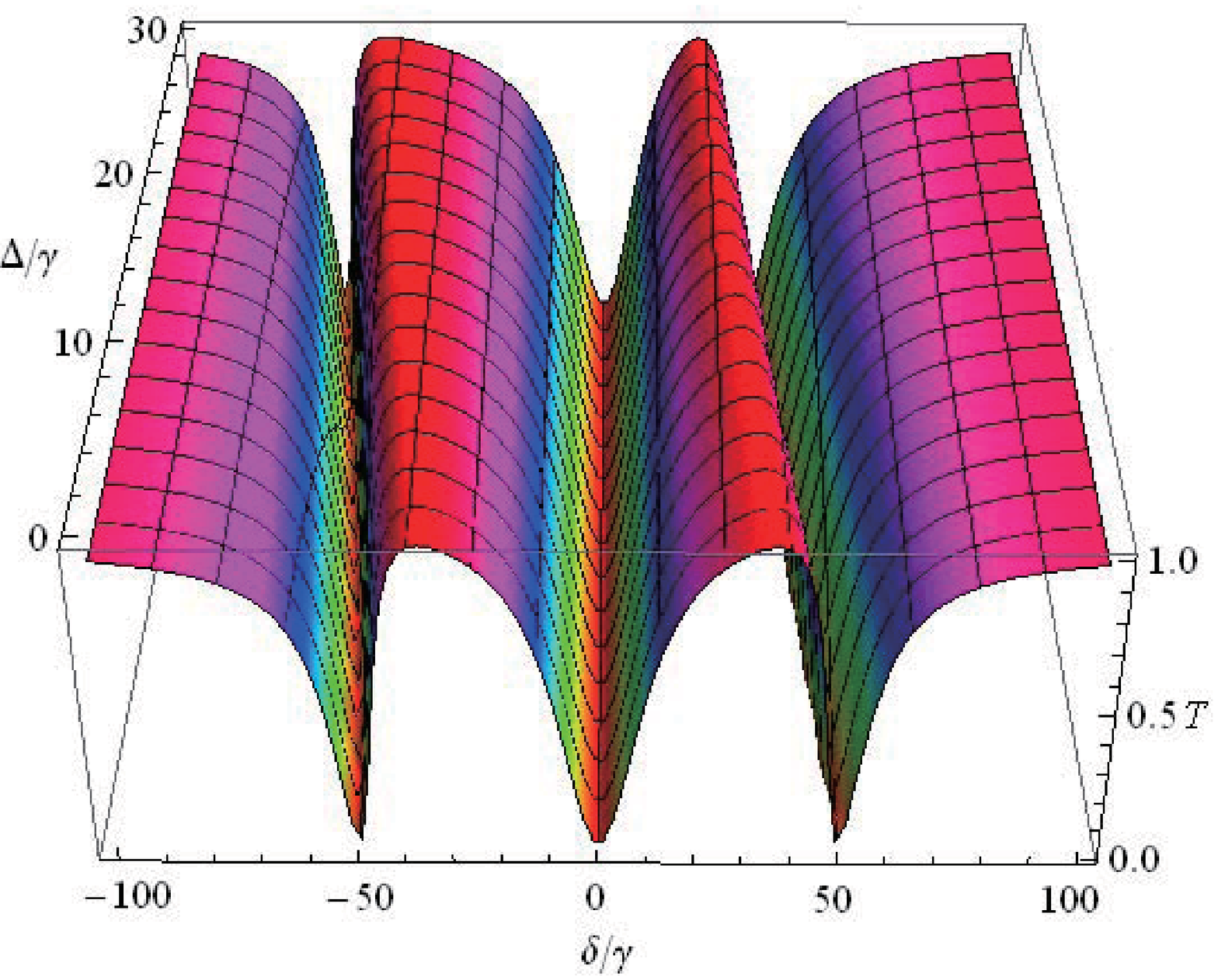}}
\end{center}
\caption{The influence of various parameters on the transmission probability. (a) The influence of the classical driving on the transmission spectrum. (b) The influence of the detuning $\Delta$ on the transmission spectrum. Here, the energies $\omega$, coupling constant $\Omega$, detuning $\Delta$ and the coupling strength $\Gamma_{i}(i=1,2)$ are in unit of $\gamma$. The parameters are set as $\Gamma_1=\Gamma_2=10\gamma$, $\Delta=0$(in fig.2(a)), $\gamma_2=\gamma_3=\gamma_4=\gamma$.}
\end{figure}

Now, suppose that the atom is originally prepared at its ground state and the right-traveling photons are sent from the far left. After the interaction with the atom, the injected single photon may be absorbed by the atom or scattered into the left or the right direction. Therefore, the general state of the whole system is
\begin{eqnarray}
\nonumber|\psi\rangle &=&\int dx \left[\phi_{R}(x)C_{R}^\dag(x)+\phi_{L}(x)C_{L}^\dag(x)\right]a^\dag_1|0\rangle\\
&+&f_2a^\dag_2|0\rangle+f_3a^\dag_3|0\rangle+f_4a^\dag_4|0\rangle,
\end{eqnarray}
where $f_2$ ($f_3$,$f_4$) is the probability amplitude of the atom in the $|2\rangle$ ($|3\rangle$,$|4\rangle$). After the interaction with the atom, one could observe the reflected or transmitted wave. The corresponding scattering amplitudes $\phi_{R(L)}$ take the form
\begin{eqnarray}
\nonumber  \phi_{R} &=& e^{ikx}\left[\theta(-x)+t\theta(x)\right], \\
  \phi_{L} &=& re^{-ikx}\theta(-x),
\end{eqnarray}
where $\theta(x)$ is the unit step function, which equals unity when $x>0$ and zero when $x<0$. Moreover, $t$ and $r$ are the transmission and reflection amplitudes in the waveguide, respectively. By solving the eigenvalue equation $H|\psi\rangle=\omega|\psi\rangle$, one can analytically obtain transmission and reflection coefficients,
\begin{widetext}
\begin{equation}\label{tran}
t=\frac{\left(\omega_3-\omega-i\frac{\gamma_3}{2}\right)\left(\omega_2-\omega-i\frac{\gamma_2}{2}
-\frac{\Omega^2}{\omega_2-\omega+\Delta-i\frac{\gamma_4}{2}}\right)}{\left(\omega_3-\omega
-i\frac{\gamma_3}{2}-i\Gamma_2\right)\left(\omega_2-\omega-i\frac{\gamma_2}{2}-i\Gamma_1
-\frac{\Omega^2}{\omega_2-\omega+\Delta-i\frac{\gamma_4}{2}}\right)+\Gamma_1\Gamma_2},
\end{equation}
and
\begin{equation}
r=\frac{i\Gamma_1\left(\omega_3-\omega-i\frac{\gamma_3}{2}\right)+i\Gamma_2\left(\omega_2-\omega-i\frac{\gamma_2}{2}
-\frac{\Omega^2}{\omega_2-\omega+\Delta-i\frac{\gamma_4}{2}}\right)}{\left(\omega_3-\omega
-i\frac{\gamma_3}{2}-i\Gamma_2\right)\left(\omega_2-\omega-i\frac{\gamma_2}{2}-i\Gamma_1
-\frac{\Omega^2}{\omega_2-\omega+\Delta-i\frac{\gamma_4}{2}}\right)+\Gamma_1\Gamma_2},
\end{equation}
\end{widetext}
of single photons transporting along the waveguide.
Above, $\Gamma_i=V_i^2/v_g$ ($i=1,2$).
\begin{figure}[b]
\begin{center}
\subfigure[]{\includegraphics[width=0.33\textwidth]{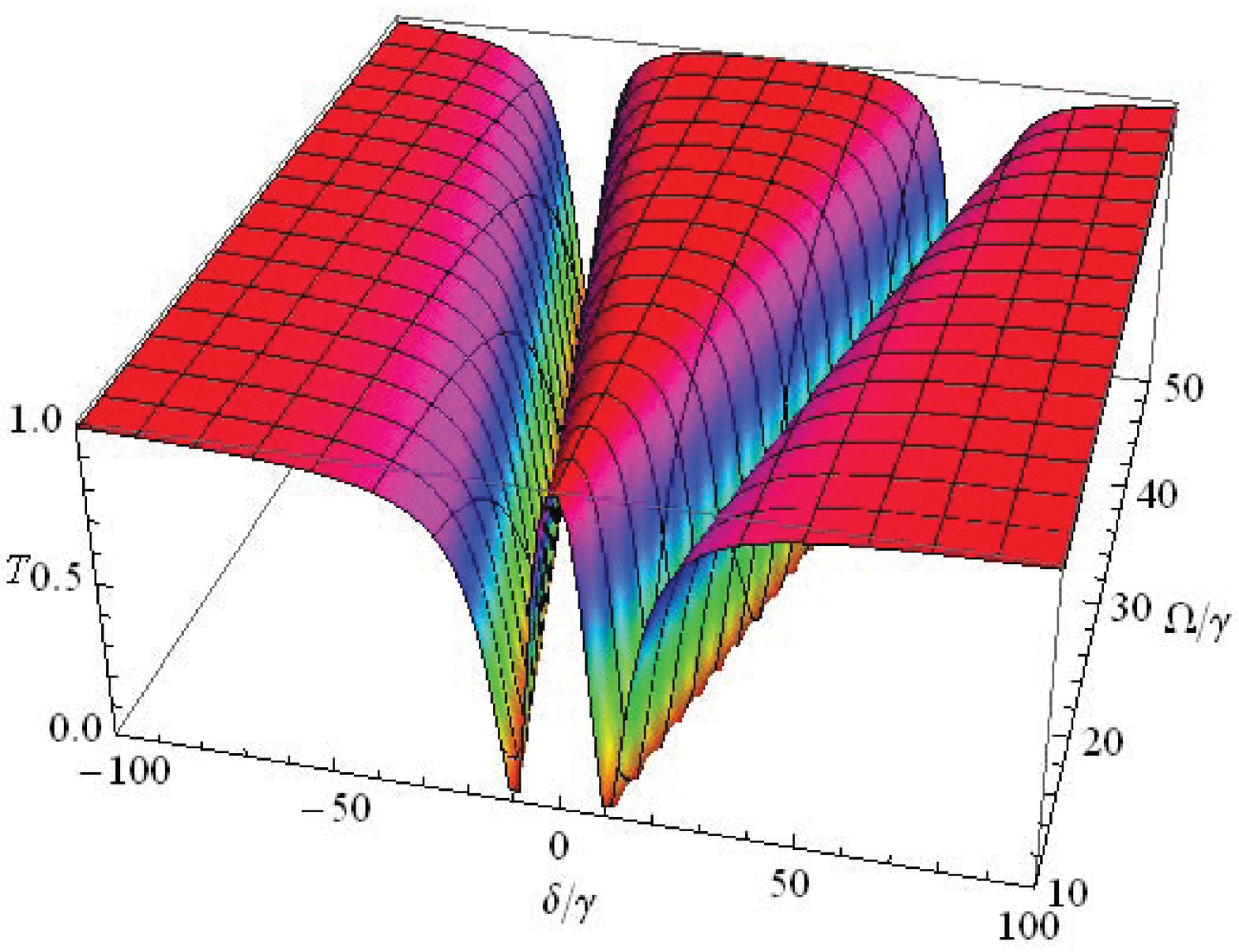}}
\subfigure[]{\includegraphics[width=0.33\textwidth]{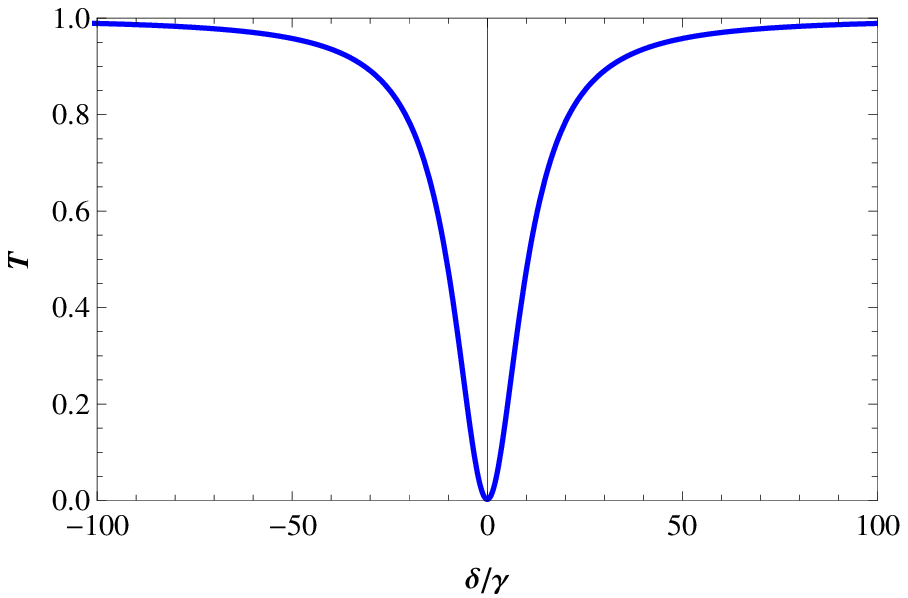}}
\end{center}
\caption{The transmission spectrum of the left or right-circularly polarized photon ($a$) or ($b$) without dissipation. The parameters are set as the ones in fig.2}
\end{figure}

%\emph{Transmission spectra of the incident single photon.--}
We now analyze numerically the transmission properties of the single photons transporting along the waveguide with a four-level atom under the classical driving $\Omega$. In Fig.~2(a), the transmission probability versus detuning $\delta$ and $\Omega$ is plotted, where $\delta=\omega_2-\omega$. For simplicity, we consider the degenerate case, namely, $\omega_2=\omega_3$. It is clearly that if the incident photon is resonance with the atomic transition $|1\rangle\leftrightarrow|2\rangle$, then the photon is completely reflected. It resembles the case with a two-level atom as in Ref.~\cite{3}, and is completely differ from the results of three-level atom shown in Ref.~\cite{6}. In Ref.~\cite{6}, when the incident photon is resonance with the atom, the photon can transmit along the waveguide freely. Moreover, in our model when $\delta=\pm \Omega$ the transporting photon is completely reflected.

An interesting feature is, the classical driving $\Omega$ influence obviously the transmission spectrum. As in Fig.~2(a), when the classical driving is applied and the coupling strength $\Omega$ increases slowly from $0$ to $20\gamma$, two peaks appear. For example, if no external driving, the transmission probability for the atom with energy $\omega=\omega_2+10\gamma$ is $50$\%. Due to the influence of the classical driving, its probability decreases to zero when $\Omega=10\gamma$.

Also, detuning parameter $\Delta$ is another important factor affecting the transmission probability of the incident photon. In Fig.~2(b), we plot the transmission spectrum versus the variation $\delta$ and $\Delta$. It is shown that the position of perfect reflection at $\delta=0$ remains unchanged, while the position of perfect reflection at $\delta=\pm \Omega$ is shifted due to the influence of the detuning $\Delta$.

Our results demonstrated above showed that, the quantum transport of single photons along an one-dimensional waveguide can be controlled by applying the classical driving to another transition path of the atom. This provides a convenient way to design certain single-quantum devices with controlled single atoms.

%\emph{``Malus's law'' at single-photon level.--}
A direct application of the above generic proposal is to control the polarization-dependent transport of the incident photon. To this end let's suppose that the transitions $|1\rangle\leftrightarrow|2\rangle\,(|3\rangle)$ are coupled to the left-circularly (right-circularly) polarized mode of photon with the coupling strength $V_1$\, ($V_2$). When a photon with left-circularly polarized mode is incident from the left of the waveguide, only the energy levels $|1\rangle$, $|2\rangle$, $|4\rangle$ of the atom are involved which resembles the case of the three level in Ref.~\cite{6}.
In this case the transmission amplitudes of photon in Eq.~(\ref{tran}) is reduced to
\begin{equation}
t_L=\frac{\omega_2-\omega-i\frac{\gamma_2}{2}-\frac{\Omega^2}{\omega_2-\omega+\Delta-i\frac{\gamma_4}{2}}}{\omega_2-\omega-i\frac{\gamma_2}{2}-\frac{\Omega^2}{\omega_2-\omega+\Delta-i\frac{\gamma_4}{2}}-i\Gamma_1}
\end{equation}
\begin{figure}[t]
\begin{center}
\subfigure[]{\includegraphics[width=0.35\textwidth]{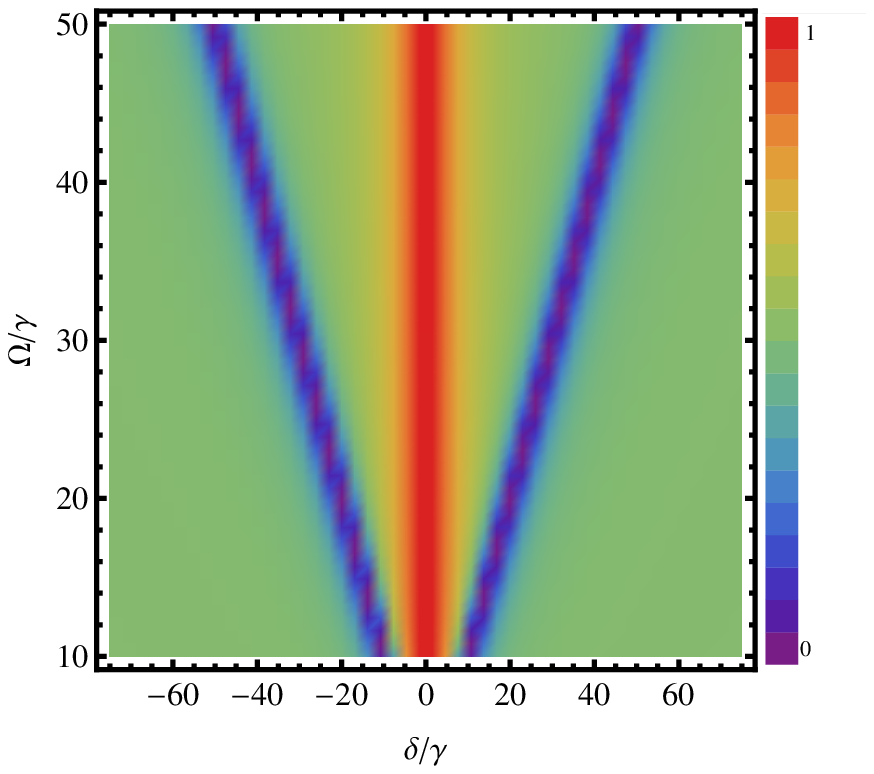}}
\subfigure[]{\includegraphics[width=0.35\textwidth]{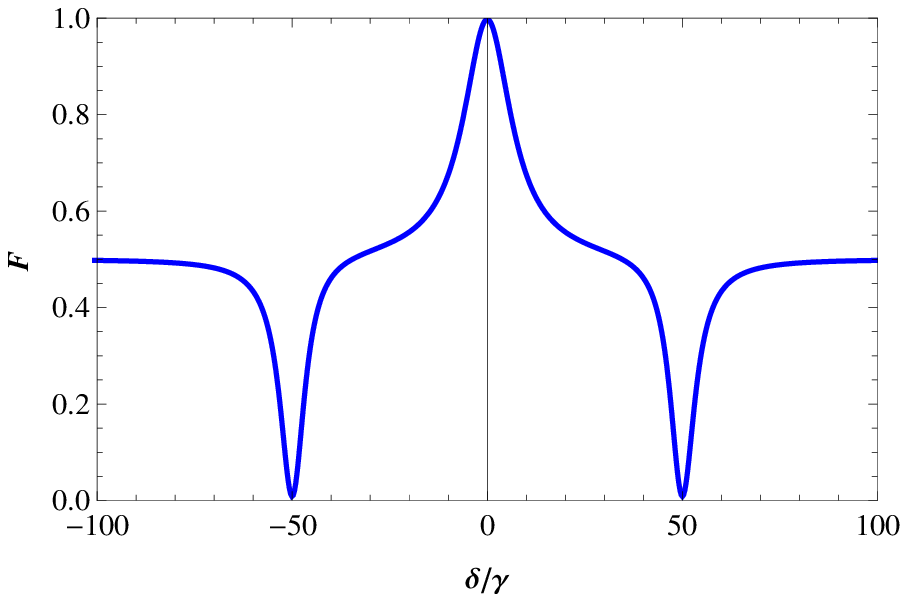}}
\end{center}
\caption{The influence of coupling strength $\Omega$ on the fidelity $F$. We set $\Omega=50\gamma$ in fig.4(b)}
\end{figure}
Therefore, if the energy of the incident photon is resonant with the transition between $|1\rangle$ and $|2\rangle$, the photon can transport along the waveguide freely. As shown in Fig.~3(a), the width of the transparency ``window'' can be adjusted by the coupling strength $\Omega$ of the external driving.
On the other hand, when a right-circularly polarized photon travels along the waveguide, it only couples the transition $|1\rangle\leftrightarrow|3\rangle$ of the atom. The transmission amplitude now is
\begin{equation}
t_R=\frac{\omega_3-\omega-i\frac{\gamma_3}{2}}{\omega_3-\omega-i\frac{\gamma_3}{2}-i\Gamma_2}.
\end{equation}
One can see from Fig.~3(b) that, if the incident photon is resonantly coupled with the atomic transition between $|1\rangle$ and $|3\rangle$, it will be reflected by the atom completely. Assume that level $|2\rangle$ and $|3\rangle$ of the atom are nearly degenerate, and the classical driving is strong enough, e.g., $\Omega=50\gamma$ then the width of transparency ``window'' is significantly large.

Now let a beam of photons with energy $\omega_3$ and unknown polarization travel from the left of the waveguide
shown in Fig.~1. The present problem is, if the photon can transport through the waveguide with a driven atom?
Generally, the state of the photon with unknown polarization can be described as
\begin{equation}
|\phi\rangle=\cos\alpha|L\rangle+\sin\alpha|R\rangle.
\end{equation}
Based on the above discussion, right-circularly polarized photons are reflected completely and the left-circularly polarized ones can be transmitted freely through the waveguide. Since the probability of the incident photon with the $L$-polarization (i.e., at the left-circularly polarized state $|L\rangle$) is $P_L=\cos^2\alpha$, the photon transmit through the waveguide with the probability $P_L=\cos^2\alpha$. This is the so-called ``Malus's law'' at single-photon level. It is noted that {\it the present $\cos\alpha$ is not referred to the relative angle between the incoming wave and the orientation of the polarizer in the classical Malus's law}. Instead, it is the weight factor of the relevant photon polarization in the state of incident photon.

In order to check the performance of our device, we define a fidelity function: $F=|t_L|^2/(|t_L|^2+|t_R|^2)$. When $F=1$, the left-circularly polarized photons transmit freely and the right-circularly polarized ones are reflected completely. Inversely, when $F=0$, the left-circularly polarized photons are blocked.
In the present single-quantum device, the imperfect factor originates mainly from the dissipation of the embedded atom such as the spontaneous emissions of the atomic excited states. Indeed, we numerically investigate these influences on the fidelity in Fig.~4(a), the fidelity as the functions of the parameters $\Omega/\gamma$ and $\delta/\gamma$. It is shown that a reliable quantum device with high fidelity (e.g., larger than 99.7\%) can be achieved at $\delta=0$ in the presence of the atomic spontaneous decays.

%\emph{Conclusions.--}
In summary, we propose a ``single-atom polarizer'', which is form by a one dimensional waveguide with a driven four-level atom. Through theoretically investigating single-photon scattering, we found that the transmission can be controlled by the selective classical driving. By adjusting coupling strength and detuning of the external driving, we show that the single photons can be transmitted or reflected, depending on the selection of the additional transition paths of the atom. Importantly, our model can be applied to realize polarization filtering at single-photon level. Meanwhile, it showed a Malus' law-like expression in the photon transport within one dimension waveguide. Accordingly, one can control the single polarized-photon transport, either transmit or reflect along the waveguide with a four-level atom by using a classical beam to selectively excite the additional transition paths of the atom. This so-called single-atom polarizer suggests that an ideal polarized-filtering device can be developed for the single-photon transistors and quantum information processing based single-photon technique.

\section*{Acknowledgments}

We are very grateful to Dr. Dai-he Fan for providing useful references. This work was supported in part by the National Natural Science Foundation of China, under Grants No. 90921010 and No. 11174373, and the National Fundamental Research Program of China, through Grant No. 2010CB923104.

%\begin{thebibliography}{000} %for 3 digits
%\begin{thebibliography}{00}  %for 2 digits

\end{document}